\documentclass[prl,reprint,superscriptaddress,nobalancelastpage,twocolumn,showpacs]{revtex4-1}

\usepackage{color,amsthm,amsmath,amsfonts,graphicx,bm,amssymb,hyperref}
\usepackage{epstopdf}
\usepackage{comment}
\usepackage{etoolbox}
\usepackage{tikz}
\usetikzlibrary{shapes,snakes}

\newcommand{\beq}{\begin{equation}}
\newcommand{\eeq}{\end{equation}}

\def\uu{u}
\def\ww{w}
\def\curr{\mathcal{J}}
\def\mm{m}
\def\kk{{\mathcal{K}}}

\newcommand{\avg}[2]{\langle #1 \rangle_{\mbox{\tiny #2}}}
\def\Tr{\operatorname{Tr}}

\def\rhostat{\rho_{\mbox{\tiny NESS}}}

\newcommand{\avgness}[1]{\avg{#1}{NESS}}

\newcommand{\titleinfo}{Energy transport in Heisenberg chains beyond the Luttinger liquid paradigm} 
\begin{document}

\title{\titleinfo} 

\author{Andrea De Luca}
\email{andrea.deluca@lpt.ens.fr}
\affiliation{Laboratoire de Physique Th\'eorique de l'ENS \& Institut de Physique Theorique Philippe Meyer, Paris, France.}

\author{Jacopo Viti}
 \email{jacopo.viti@lpt.ens.fr}
\affiliation{Laboratoire de Physique Th\'eorique de l'ENS, CNRS \& Ecole Normale Sup\'erieure de Paris, Paris, France.}

\author{Leonardo Mazza}
 \email{leonardo.mazza@sns.it}
\affiliation{NEST, Scuola Normale Superiore \& Istituto Nanoscienze-CNR, Pisa, Italy. }

\author{Davide Rossini}
 \email{davide.rossini@sns.it}
\affiliation{NEST, Scuola Normale Superiore \& Istituto Nanoscienze-CNR, Pisa, Italy. }

\date{\today}

\begin{abstract}
We study the energy transport between two interacting spin chains
which are initially separated, held at different temperatures 
and subsequently put in contact.
We consider the spin-1/2 XXZ model in the gapless regime 
and exploit its integrability
properties to formulate an analytical Ansatz for the non-equilibrium 
steady state even at temperatures where the low-energy Luttinger liquid
description is not accurate.
We apply our method to compute the steady energy current and benchmark it
both with the known low-energy limit and at higher temperatures 
with numerical simulations. 
We find an excellent agreement even at high temperatures, where the 
Luttinger liquid 
prediction is shown to fail.
\end{abstract}


\maketitle

\paragraph{Introduction. --- }

The study of heat propagation is a fertile research field 
in condensed matter physics~\cite{blanter2000shot}.
Up to few years ago, low-dimensional transport experiments 
have been always considered a prerogative of solid state nanowires. 
Any measurable current is modeled as a flow between 
incoherent and non-interacting reservoirs~\cite{Kane1996, *Fazio1998},
so that transport phenomena in this scenario are well described 
within the Landauer-B\"uttiker approach~\cite{datta1997electronic}.
However, the latest groundbreaking advances with cold atoms have challenged this paradigm:  
thermoelectric transport can now be studied with high degree of control and tunability 
in globally closed systems, where the interplay between interactions and coherence 
is potentially crucial~\cite{gring2012relaxation, *Brantut2013, *Krinner2014}.

\begin{figure}[t]
  \includegraphics[width=1.0\columnwidth]{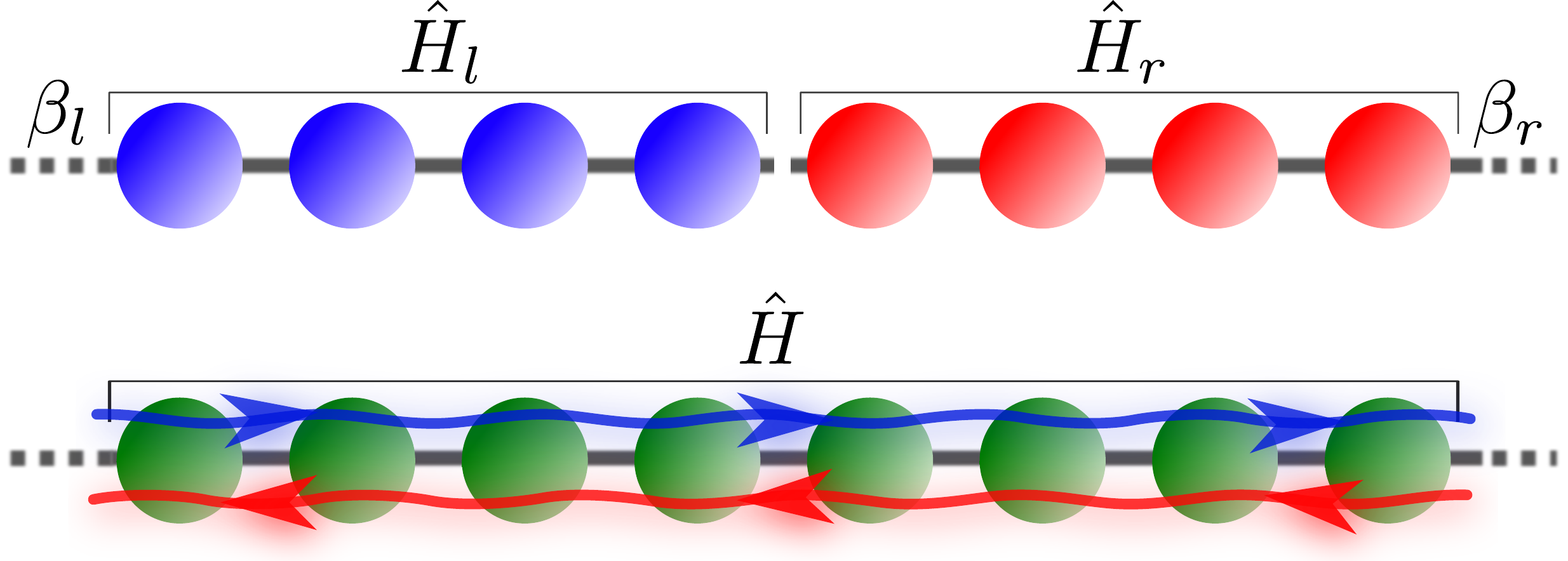}
  \caption{Sketch of the non-equilibrium protocol:
    Two initially disconnected half chains of semi-infinite length, 
    thermalized at different inverse temperatures $\beta_l$ and $\beta_r$, 
    are connected at time $t_0 \equiv 0$. 
    A net energy current $\mathcal{J}$ flowing through the system 
    is expected to appear once the steady state is established. 
    At low temperature, this is interpreted as the effect of separate thermalization of the right/left moving excitations (lines with arrows).
    }
  \label{fig:Sketch}
\end{figure}

A recent theoretical work~\cite{bernard2012energy, *bernard2013non} investigated two semi-infinite 1D reservoirs 
described by a conformal field theory (CFT)
and held at different temperatures~\cite{spohn1977stationary}. 
At a certain time $t_0$, they are directly put in contact restoring translational invariance and evolved unitarily (see Fig.~\ref{fig:Sketch}).
There it is shown the existence of a non-equilibrium steady state (NESS) 
featuring an energy current that only depends 
on the central charge $c$ specifying the CFT of the reservoirs. 
CFT~\cite{mathieu1997conformal} describes the low-energy physics of 1D gapless systems and a standard framework to characterize electronic wires at low temperatures is
the Luttinger Liquid (LL)~\cite{haldane1981luttinger, *haldane1981effective},  a CFT with $c=1$.
It follows that the thermal current between 1D electron reservoirs has a universal low-temperature behavior. 
The presence of a persistent current and its universality have been rigorously 
proven in free models~\cite{ho2000asymptotic,*tasaki2001nonequilibrium, *aschbacher2003non, de2013nonequilibrium} 
and numerically verified in the scaling limit of critical spin chains~\cite{karrasch2012nonequilibrium}.

However, the LL model describes the low-energy excitations of the interacting 
system~\cite{giamarchi2004quantum}, and its predictions are deemed to fail 
far from equilibrium and whenever too high temperatures are considered. 
Various numerical~\cite{HeidrichMeisner2003, HeidrichMeisner2007, karrasch2012finite, *Karrasch2013, karrasch2013drude, karrasch2014real, bonnes2014light} 
and analytical studies~\cite{Zotos1997, Jung2006, sirker2009diffusion, *sirker2011conservation, prosen2011open, *prosen2013families, *prosen2014lower} 
have attacked the problem, highlighting the peculiar effects of integrability.
Moreover, recent experimental results~\cite{sologubenko2007thermal, hess2007heat, hlubek2010ballistic} 
provided evidence of ballistic heat transport of quantum spin excitations. 

In this paper we study two reservoirs described by
the XXZ spin-1/2 Hamiltonian in the gapless regime. 
This model is unitarily related to a system of interacting 
spinless fermions~\cite{giamarchi2004quantum}.
Its special integrability properties suggest the existence of a NESS energy current 
even at high temperatures~\cite{klumper2002thermal, karrasch2012nonequilibrium}. An analytical treatment based on the
thermodynamic Bethe Ansatz has been put forward
for the isotropic ferromagnet exploiting the low density of magnons at low-temperature~\cite{karrasch2012nonequilibrium}.
As expected, this approximation is not appropriate to reproduce the conformal regime considered here.
We discuss an Ansatz for the NESS density matrix 
based on the quantum-transfer-matrix (QTM)
formalism~\cite{klumper1993thermodynamics, klumper2000thermodynamics, klumper2004integrability}.
We propose an analytical method to characterize the thermal current 
at arbitrary temperatures. 
In the low-temperature limit, our approach reproduces 
exactly the LL prediction~\cite{bernard2012energy, *bernard2013non}.
Upon comparison with matrix-product-state (MPS) 
simulations~\cite{karrasch2012finite, *Karrasch2013, Schollwoeck2011}, 
the method is shown to be almost exact even at high temperature.
In the latter regime, it discloses distinctive signatures of interactions
in the energy transport of 1D fermionic systems.

\paragraph{The Model. ---}
\label{sec_ness}

We consider the Hamiltonian
\begin{align}
  & \hat H = \hat H_l + \hat H_r + \hat h_0 \;, \quad
  \hat H_{l} = \sum_{n < 0} \hat h_n \; , \quad
  \hat H_{r} = \sum_{n > 0} \hat h_n \; , \nonumber
  \\
  &\hat h_n = J \bigl[\hat S_{n}^{x} \hat S_{n+1}^{x} +
    \hat S_{n}^{y} \hat S_{n+1}^{y}
    +\Delta \hat S_{n}^{z}\hat S_{n+1}^{z} \bigr]\; ,
  \label{XXZhamiltonian}
\end{align}
$\hat S_n^{\alpha}$ being the $\alpha$-th component of the spin-$1/2$ 
on site $n$ ($\hbar = k_{\rm B} = 1$); we focus on the  
critical phase $-1 < \Delta \leq 1$, 
parameterizing $\Delta = \cos\gamma$, $\gamma \in [0, \pi)$. Numerical results will be given
for the antiferromagnetic regime $\gamma \in [0, \pi/2]$. 
At the beginning, the system is separated into 
two independent halves held at different inverse temperatures $\beta_{r,l}$ 
\beq
  \label{rhozero}
  \rho_0 = \mathcal Z^{-1} e^{- \beta_l \hat{H}_l}\otimes e^{ - \beta_r \hat{H}_r} \;
\eeq
where $\mathcal Z$ ensures normalization of $\rho_0$.
For times $t \geq 0$ the state is unitarily evolved 
with Hamiltonian $\hat H$~\eqref{XXZhamiltonian}, so that the initially-separated
reservoirs are put in contact and generate a heat flow (see Fig.~\ref{fig:Sketch}).

The existence of a NESS requires the convergence of the long-time limit
\beq
  \label{rhostatdef}
  \lim_{t\to\infty}\lim_{N\to\infty} \Tr \big[
    \hat {\mathcal O} e^{- i \hat H t} \rho_0 e^{i \hat H t}
    \big] \equiv \avgness{\hat {\mathcal O}}
\eeq
for any local observable $\hat {\mathcal O}$, where $N$ is the system size. 
Equation~\eqref{rhostatdef} formally defines a density matrix $\rhostat$ 
embodying all the local properties of the long-time dynamics.

In some peculiar situations the structure of $\rhostat$ can be inferred from 
general symmetry arguments, as translational invariance. 
Suppose that the excitations of the system can be separated 
into two non-interacting sets with positive ($+$) and negative ($-$) momenta, 
so that $\hat H = \hat H_+ + \hat H_-$ and $[\hat{H}_{+},\hat{H}_{-}]=0$
(here by positive we mean going from left to right, and vice-versa---see Fig.~\ref{fig:Sketch}).
Then $\rhostat = \mathcal Z^{-1} e^{- \beta_l \hat H_+} \otimes e^{- \beta_r \hat{H}_-}$, 
{\it i.e.}, the right (left) movers with positive
(negative) momenta are separately thermalized  at $\beta_{l}$ ($\beta_r$). 
This happens notably in any CFT, where
the expectation value of the energy current operator in the middle of the chain
$\hat J_E = (i/2)[ \hat H, \hat H_r - \hat H_l]$ 
is~\cite{bernard2012energy, *bernard2013non} 
\beq
  \label{currconf}
\curr = \avgness{\hat J_E} 
  = \frac{\pi c}{12}  \left( \, \beta_l^{-2} - \beta_r^{-2} \, \right) \,,
\eeq
with $c=1$ for the critical XXZ model. 
However when the temperatures are increased, 
corrections appear due to irrelevant operators~\cite{lukyanov1998low} 
that couple the right and left movers and spoil the pure conformal result~\eqref{currconf}. 

\paragraph{The QTM approach. ---}
We now explain the method to extend Eq.~\eqref{currconf} to higher temperatures 
for the XXZ model. First we briefly recall how the QTM formalism allows to extract 
thermodynamical quantities at equilibrium, {\it i.e.} when $\beta_{l}=\beta_{r}=\beta$. 
Excitations of this spin chain can be interpreted as quasiparticles 
and quasiholes with energy~\cite{vsamaj2013introduction} 
\beq
  \label{energy_op}
  \varepsilon(\uu) = \frac{\pi \Lambda  }{2\cosh(\pi \uu/2)}\;,
\eeq
where $\Lambda = 2 J \sin\gamma / \gamma$ and $u \in (-\infty, \infty)$ parametrizes 
the first Brillouin zone, such that $u>0$ ($u<0$) corresponds to positive 
(negative) momenta. Thermodynamic quantities can be obtained knowing
the quasiparticles and quasiholes occupation numbers $n(u)$ and $\bar n(u)$ at finite temperature.
It is useful to rewrite $n\equiv b/(1+b)$ (and analogously for $\bar n$), where $b(\uu)$ and $\bar b(\uu)$ 
solve the non-linear integral equation~\cite{klumper2000thermodynamics}
\begin{align}
  \label{NLIE}
  \left[\begin{array}{c}
      \log b\\
      \log\bar{b}
    \end{array}\right]=\left[\begin{array}{c}
      s\\
      s
    \end{array}\right]+\left[\begin{array}{cc}
      \kk& -\kk_+ \\
      -\kk_-& \kk
    \end{array}\right]\star\left[\begin{array}{c}
      \log (1+b)\\
      \log(1+\bar{b})
    \end{array}\right],
\end{align}
where $f\star g$ is the usual convolution. 
At equilibrium the source therm $s(\uu)$ in~\eqref{NLIE} is $s(\uu)=-\beta\varepsilon(\uu)$. 
The kernel $\kk(u-v)$ is the logarithmic derivative of the scattering phase
between two quasi-particles or quasi-holes at different momenta. Explicitly one has
\beq
  \label{kernel}
  \kk(\uu)=\int_{-\infty}^{\infty}\frac{d\ww}{2\pi}~\frac{\sinh\left[\left(\frac{\pi}{\gamma}-2\right)\ww\right]}
     {2\cosh \ww\sinh\left[\left(\frac{\pi}{\gamma}-1\right)\ww\right]}e^{i\ww\uu};
\eeq
and $\kk_{\pm}(\uu)=\kk(\uu\pm 2i)$. 
Notice that, for $\Delta=0$ ($\gamma=\pi/2$), $\kk$ vanishes and Eq.~\eqref{NLIE} 
reduces to the usual Fermi-Dirac distribution $n=1/(1+\text{exp}(\beta\varepsilon))$ 
for free fermions. The advantage of using the QTM approach and Eq.~\eqref{NLIE} 
lies in its low-temperature limit $\beta\rightarrow\infty$. 
As we will discuss below, at low temperatures 
Eq.~\eqref{NLIE} decouples into two \textit{independent} equations
determining the occupation numbers for excitations with positive ($u>0$) and negative momenta ($u<0$).
In the $\Delta\rightarrow 0$ limit they correspond to the two chiral components of
a free Dirac fermion.

In order to describe this out-of-equilibrium protocol, we propose to use Eq.~\eqref{NLIE} 
assuming a source term of the form $s(\uu) = -\beta(\uu) \varepsilon(\uu)$ 
with $\beta(\uu) = \beta_{r}$ ($\beta_l$) for $\uu<0$ ($\uu>0$).
Indeed, in this setting, the quasi-particles and quasi-holes can be supposed to be thermalized
at large distances from the center of the system
where they are infinitely far apart and interactions are negligible.
Then, as it happens at equilibrium, bulk occupation numbers
are modified by the phase shifts collected in each scattering process, which result in the
convolution term of \eqref{NLIE}. Since this term is the effect of the microscopic
processes appearing in an integrable model, it is natural to assume that it is not affected
by the initial conditions.
Despite this hypothesis, the proper derivation of the source term would require 
the exact solution of the dynamics at long times, an extremely challenging task even for
integrable models. Our choice is exact at the free-fermion point ($\Delta = 0$)~\cite{de2013nonequilibrium}
and, as we will see, yields the correct low-temperature behavior for all $\Delta$, 
which remained elusive in the analytic approach of~\cite{karrasch2012nonequilibrium}.
The idea of thermalizing independently right-left-moving quasi-particles was already 
stated in~\cite{bernard2012energy, doyon2012nonequilibrium, *castro2014thermodynamic}.

The knowledge of $n(\uu)$ permits computing the expectation value in the NESS 
of any local operator~\cite{gohmann2004integral}, and in particular $\mathcal J$.
A complete set of conserved quantities $\hat {\mathcal{Q}}_{\mm}$ ($\mm \in \mathbb N^+$) 
in the XXZ chain can be defined iteratively~\cite{grabowski1995structure} starting
from the Hamiltonian $\hat H \equiv \hat{\mathcal{Q}}_1$
by repeated applications of the boost operator 
$\hat {\mathcal{B}} = \sum_{n}n \hat h_n$: 
$\hat {\mathcal{Q}}_{\mm+1}=i[\hat{\mathcal{B}}, \hat{\mathcal{Q}}_{\mm}]$.
Their expectation values in the NESS are extensive, {\it i.e.}, 
$\langle \hat{\mathcal{Q}}_\mm\rangle\sim N q_\mm$.
Because of translational invariance, 
the energy current $\mathcal J$ coincides with $- q_2$. 
Indeed, $\hat J_E=-(\hat j_0+\hat j_{-1})/2$, where $\hat j_{n}=i[\hat h_n, \hat h_{n-1}]$; 
by commuting  $\hat H=\sum_{n} \hat h_n$ with $\hat {\mathcal B}$ one easily realizes
that $\hat{\mathcal{Q}}_2=\sum_{n}\hat j_n$.

\begin{figure}[!t]
  \includegraphics[width=0.9\columnwidth]{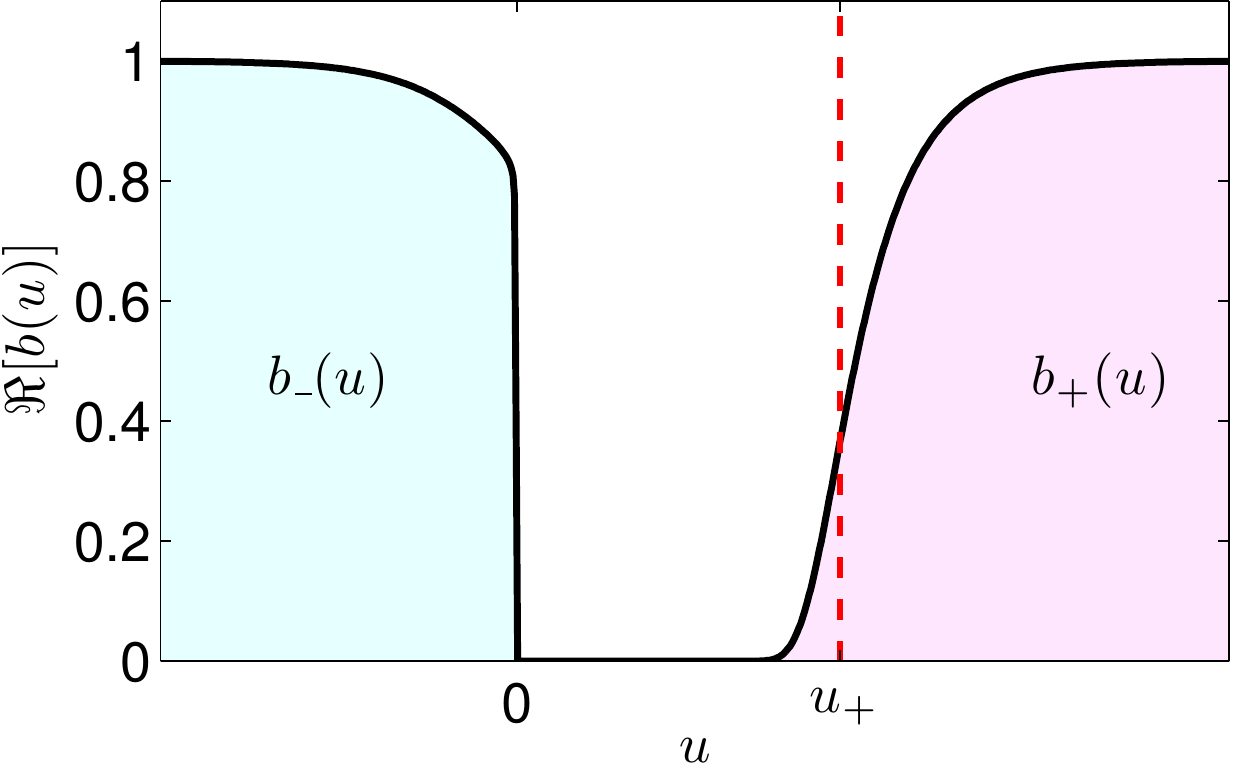}
  \caption{The real part of $b(u)$ is shown vs $u$ for $\Delta = 0.5$ 
    and $\beta_l = 10^3$, while $\beta_r = 0$. The function is exponentially suppressed for $0 <u \lesssim u_+$.
    The imaginary part similarly vanishes in the same domain.} 
  \label{plotLowT}
\end{figure}

The explicit expression for $q_{\mm}$ is
\beq
  \label{charges}
  q_{\mm}=f_{\mm}+\int_{-\infty}^{\infty}d\uu~\frac{n(\uu)a_{\mm}(\uu)+\bar{n}(\uu)\bar{a}_{\mm}(\uu)}{4\cosh (\pi \uu/2)}\,,
\eeq
where the $f_{\mm}$ are constants vanishing for even $\mm$. 
Details on the derivation of~\eqref{charges} and on its relation with the
generalized Gibbs ensemble~\cite{pozsgay2013generalized, fagotti2013stationary} 
are contained in the supplementary material.
The auxiliary functions $a_{\mm}$, $\bar{a}_{\mm}$ solve the following linear integral equation
\begin{align}
  \label{linearak}
  \left[\begin{array}{c}
      a_{\mm}\\
      \bar{a}_{\mm}
    \end{array}\right]=\left[\begin{array}{c}
      \varepsilon_{\mm}\\
      \varepsilon_{\mm}
    \end{array}\right]+\left[\begin{array}{cc}
      \kk & -\kk_+ \\
      -\kk_-& \kk
    \end{array}\right]\star\left[\begin{array}{c}
      na_{\mm}\\
      \bar{n}\bar{a}_{\mm}
    \end{array}\right]\;.
\end{align}
The function $\varepsilon_{\mm}(\uu)$ is the quasi-particle eigenvalue 
of the $\mm$-th conserved quantity, $\varepsilon_{\mm+1}(\uu)=\Lambda^{\mm}
\varepsilon^{(\mm)}(\uu)$ where $\varepsilon^{(\mm)}$ is the $\mm$-th derivative of Eq.~\eqref{energy_op}.

Eqs.~\eqref{NLIE} and \eqref{linearak} are suitable for analytical manipulations and can be easily solved numerically for $0\leq \Delta \leq 1$ by iteration, treating convolutions in Fourier space.
However, in the ferromagnetic regime, the numerical solution by iteration is problematic,
as it happens already at equilibrium~\cite{klumper2002thermal}.

\paragraph{Low-temperature limit. ---}

We first validate our Ansatz considering
the limit $\beta_l \to \infty$ while $\beta_r$ remains finite.
In this case, the function $b(\uu)$ is negligible in a finite range 
of values $\uu \in [0, \uu_+]$ (see Fig.~\ref{plotLowT}).
The value $\uu_{+}=\frac{2}{\pi}\log\bigl[\pi \Lambda \beta_{l}\bigr]$ 
is estimated neglecting the second term in the r.h.s. of Eq.~\eqref{NLIE} 
and imposing $b(\uu_+) \sim 1$.
It is natural to split $b(\uu)$ into two functions by defining 
$b_{+}(\xi)\equiv b(\uu_+ + \frac{2}{\pi}\xi)$ and $b_-(\xi) = b(-\frac{2}{\pi}\xi)$ 
for positive $\xi$ and zero otherwise (respectively violet and cyan in Fig.~\ref{plotLowT}). 
For the case of interest $\beta_l \to \infty$ ($\uu_+ \to \infty$), Eq.~\eqref{NLIE} 
decouples into two separate equations for $b_\pm$, because $\kk(\uu)$ vanishes for large $\uu$. 
Moreover the equation for $b_{+}$ ($b_{-}$) depends only on $\beta_l$ ($\beta_r$) 
and consequently the expectation value $q_{\mm}$ has the simple form
\beq
  \label{factorization}
  q_{\mm}=f_{\mm}+q_{\mm}^+(\beta_l)+q_{\mm}^{-}(\beta_r)\;, 
\eeq
where $q_{\mm}^{\pm}$ can be computed as explained in the supplementary material. 
In particular, $q_{\mm}^{+}$ is exactly 
obtained generalizing the so-called dilogarithm trick~\cite{klumper2002thermal}
\beq
\label{charge_CFT}
q_{\mm}^{+}(\beta_l)=-\frac{\pi c_{\mm}}{12 \beta_l^{2}}\;,
\eeq
with $c_{\mm+2}=(-\pi\Lambda/2)^{\mm}$.
Notice that $c_{2}=1$ and the contribution to the energy 
current $\mathcal J$ due to the right moving excitations is universal.
Assuming a large $\beta_r$, one obtains a similar expression for $q_{\mm}^{-}$ and the complete relation~\eqref{currconf} is recovered. 

\begin{figure}[t]
  \includegraphics[width=0.95\columnwidth]{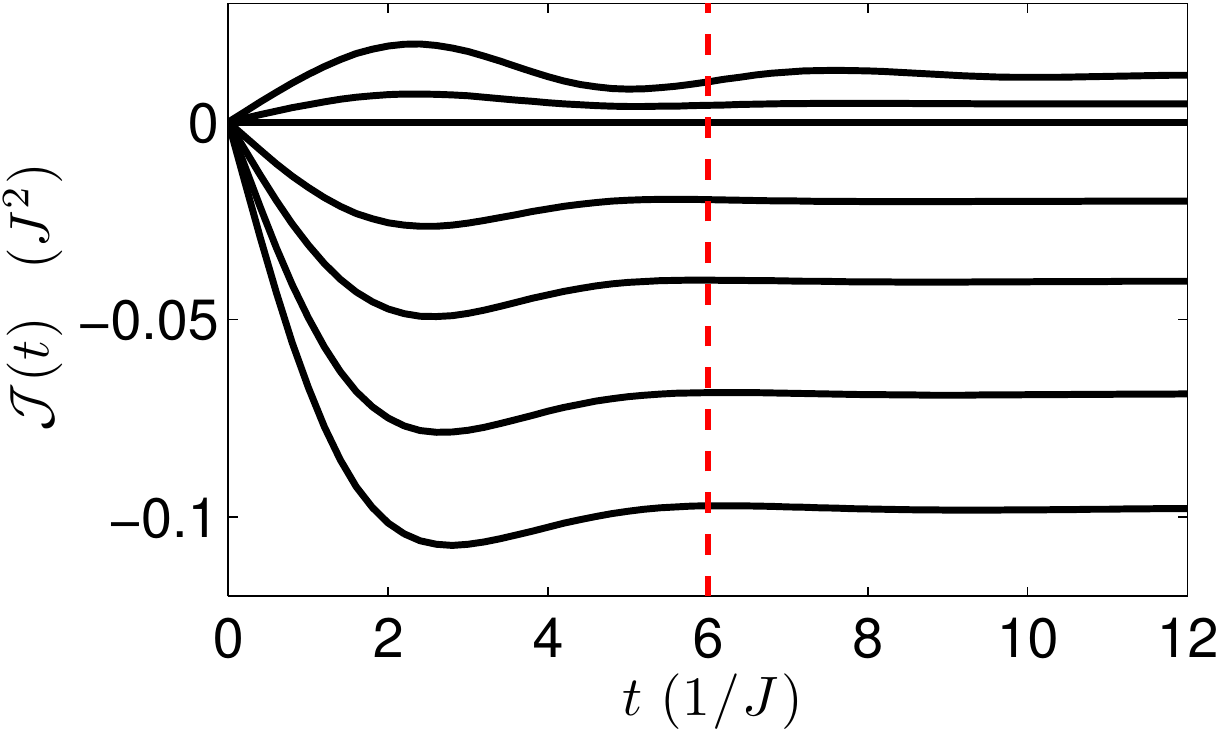}
  \caption{Time evolution of the energy current $\mathcal J(t)$ for $\Delta=0.5$
    and $\beta_l = 4 J^{-1}$.
    Different lines refer to different values of $\beta_r$,
    from top to bottom: $8.0$, $4.8$, $4$, $2.4$, $1.6$, $0.8$, $0.04~J^{-1}$.
    Steady-state values are extrapolated by averaging 
    over times $t \geq t_{\rm min}$ (see the vertical dashed line).
    Error bars in Fig.~\ref{fig:Figure4} indicate the range of values
    between the largest and the smallest 
    computed value of $\mathcal J(t)$ for $t \geq t_{\rm min}$.}
  \label{fig:Figure3}
\end{figure}

\begin{figure*}[t]
\begin{minipage}{0.495\textwidth}
\centering
\includegraphics[width=0.9\textwidth]{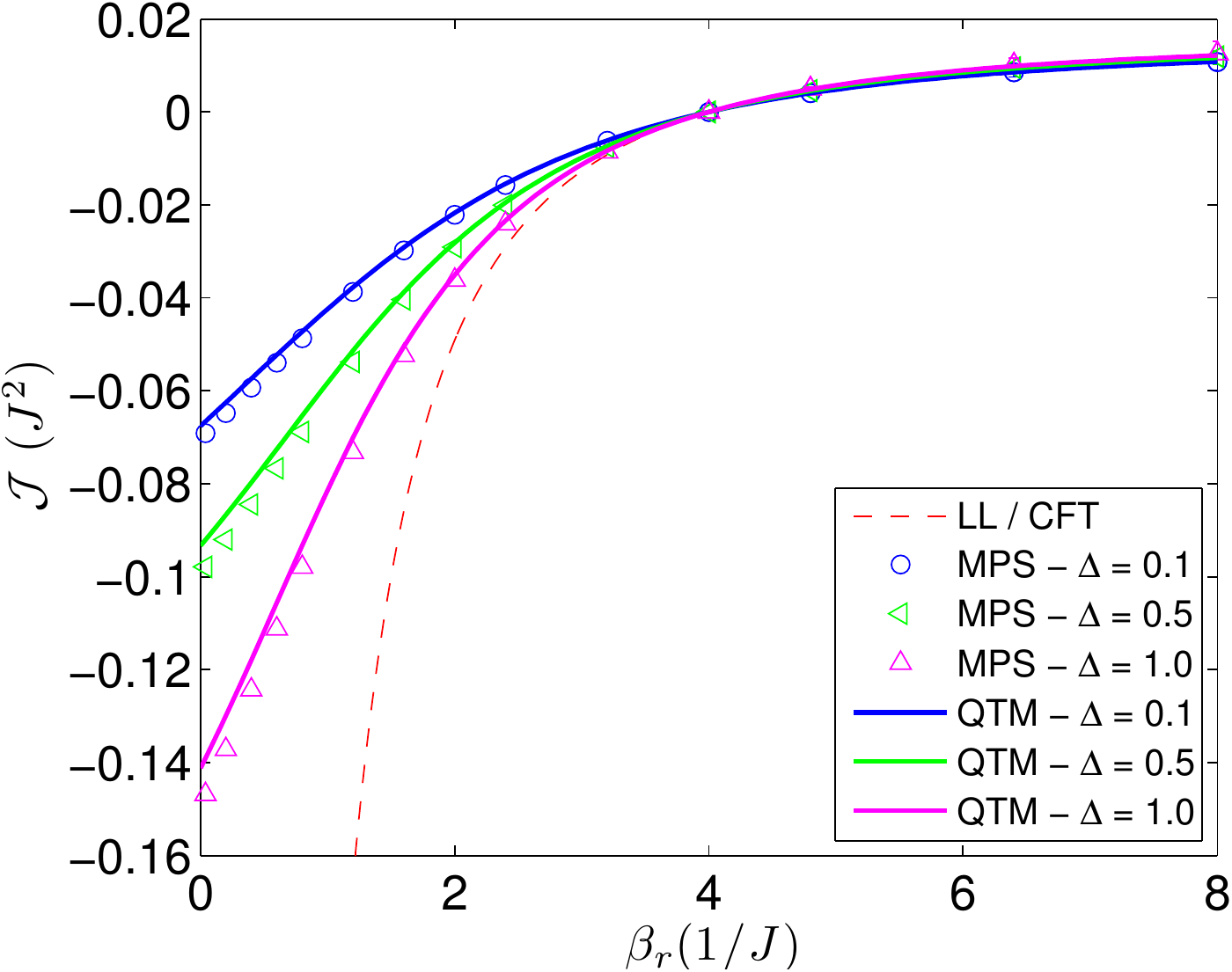}
\end{minipage}
\begin{minipage}{0.495\textwidth}
\centering
\includegraphics[width=0.9\textwidth]{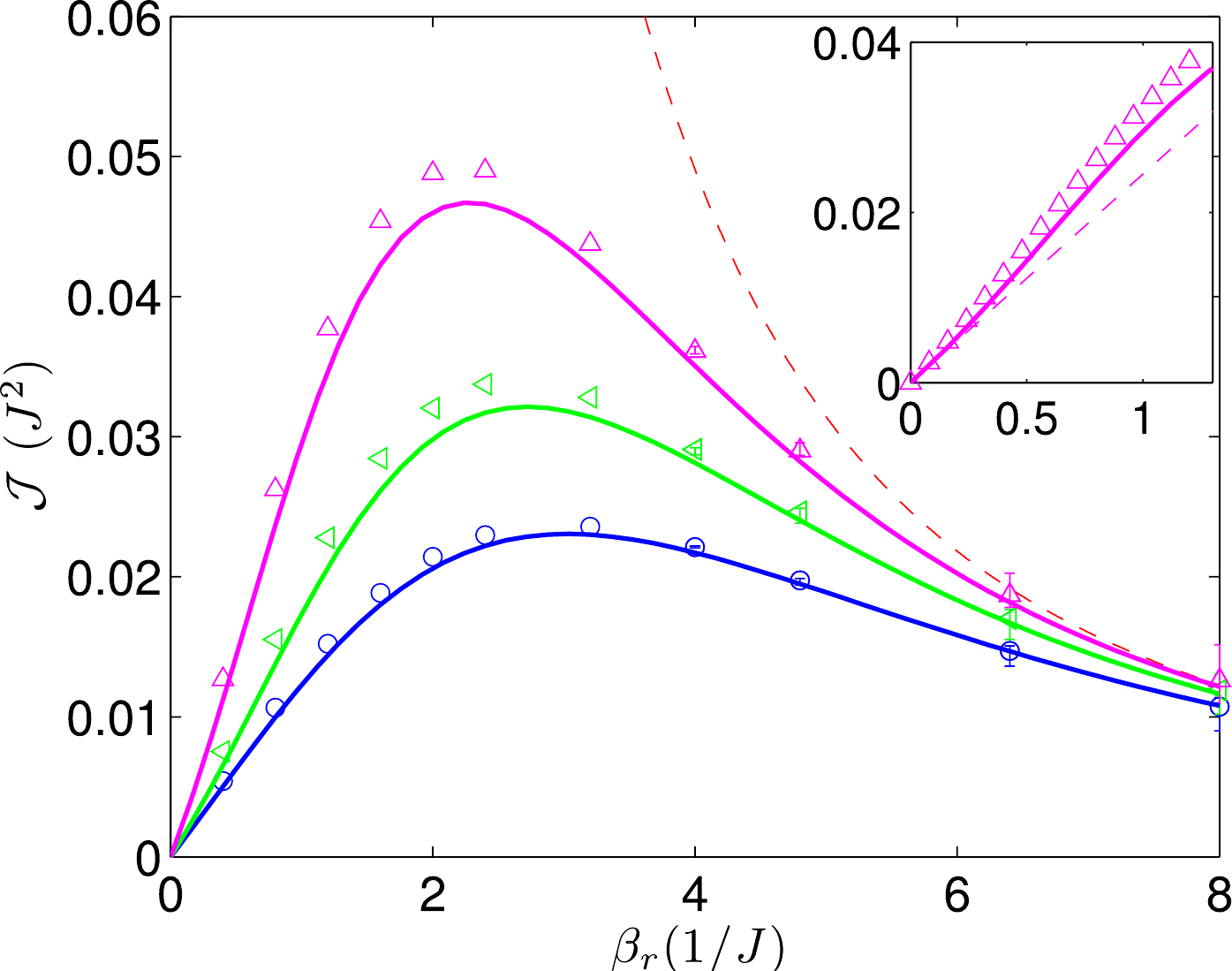}
\end{minipage}
  \caption{Energy current $\mathcal J$ for several values of $\Delta$ as a function 
    of $\beta_r$ for (a) $\beta_l = 4 J^{-1}$ and (b) $\beta_l = \beta_r/2$.
    The QTM results (continuous lines)
    are validated by the MPS data (symbols).
    When not shown, the numerical error bar is smaller 
    than the marker (see Fig.~\ref{fig:Figure3}).
    The CFT prediction~\eqref{currconf} (dashed red line) 
    describes the data {\it only} when both temperatures are small.
    The inset provides a zoom of the high-temperature. The dashed-line is the linear approximation of $\mathcal J$ for $\beta_r = 0$ (see supplementary material).
    The QTM curve shows a change of concavity which also appears in the MPS data, thus proving that our Ansatz can capture even small details of the energy current.
    }
  \label{fig:Figure4}
\end{figure*}

An interesting outcome of this approach is the splitting of the chiral degrees 
of freedom whenever one of the two temperatures is small irrespectively of the other, 
as shown in Eq.~\eqref{factorization}. 
This explains why the factorization of the current approximately holds 
for all values of the temperatures, as numerically observed in~\cite{karrasch2012nonequilibrium}.
Intuitively, $\uu_+$ can be considered as an effective Fermi point 
around which the relevant excitations are located. 
Due to the locality of the interaction, excitations with a large momentum difference do not affect each other. 
Thus, even the left moving excitation closest in momentum 
to $\uu_+$ cannot be influenced by the right moving ones.

\paragraph{Higher temperatures. ---}  \label{sec_comp}

To validate the predictions of our Ansatz at higher temperatures, we perform numerical simulations
with an algorithm based on time-dependent MPS~\cite{Schollwoeck2011}. The initial thermal state $\rho_0$ 
is computed purifying the density matrix via the ancilla method~\cite{Feiguin2005}; 
the approach for real-time evolution
of thermal states, introduced in Ref.~\cite{karrasch2012finite, *Karrasch2013, kennes2014extending}, is fundamental to reach sufficiently long times.
We consider chains up to $N=100$ with open boundary conditions;
finite-size effects are under control.
The maximum allowed bond link is $D = 3000$
and the truncation error per step is set to $10^{-10}$.
The algorithm computes the real time evolution of the density matrix, {\it i.e.} 
$\rho(t) \equiv e^{-i \hat H t} \rho_0 e^{i \hat H t}$.

In Fig.~\ref{fig:Figure3} we plot the energy current 
$\mathcal{J}(t) \equiv \text{Tr} [\hat J_E \, \rho(t)]$ 
as a function of time. 
Upon waiting enough time, the system displays clear steady signatures. 
We interpret them as distinctive features of $\rhostat$,
even if the numerics cannot guarantee the formal existence of such limit: 
longer time scales are unaccessible,
due to the exponentially increasing amount of needed resources. 
However we can estimate the steady-state limit by time-averaging $\mathcal{J}(t)$ 
from a given $t_{\rm min}$ up to the longest reached time.
Figure~\ref{fig:Figure4} compares these values 
with those derived from the analytical Ansatz. 
The agreement is excellent even in the high-temperature region, 
where the LL prediction~\eqref{currconf} completely fails, as the latter
requires both $\beta_l$ and $\beta_r$ to be large.
The QTM method is less accurate far from equilibrium
when the temperatures are of order $J$ 
(see the right panel for $\beta_l = \beta_r / 2 \sim 2 J^{-1}$), 
though here the relative inaccuracy is always found to be less than $10 \%$. 
It is difficult to address the steady state for low-temperatures
with MPS because of inaccessible long equilibration times,
thus yielding non-negligible error bars on the estimated value of $\mathcal J$.
However, in this regime the QTM method is guaranteed
to work by the presented analytical considerations.

The data confirm the intuitive expectation that the current 
is larger in situations where $\beta_l$ strongly differs from $\beta_r$. 
The non-monotonous behavior of $\mathcal J(\beta_r)$ in the 
right panel of the figure follows from the competition between 
this tendency and the fact that for $\beta_r = 0$ the system is at equilibrium.
Interestingly, in both panels we observe that
strong interactions (large values of $|\Delta|$) enhance the current.

\paragraph{Concluding remarks. ---}
We developed an analytic QTM formalism that is able to 
describe the steady-state energy current flowing between two interacting 
XXZ chains integrable through Bethe Ansatz. The method is predictive even 
far from equilibrium and at high temperatures, 
where the Landauer-B\"uttiker approach cannot be employed.

This non-equilibrium protocol can be realized in forthcoming 
cold-atom experiments~\cite{gring2012relaxation, *Brantut2013, *Krinner2014} where 
the distinctive features of the energy current predicted by our technique
could be effectively measured. The NESS is observable on a time scale of order $L/v$, where
$L$ is the system length and $v$ the typical quasi-particle velocity;
for larger times, the system will equilibrate and the current will vanish~\cite{collura2014quantum}.

Extensions to other experimentally realizable situations, 
i.e. the unbalance of chemical potentials, are possible
within the same formalism. Moreover, 
the problem of transport in the Lieb-Liniger model is under consideration by the authors.

We believe that the peculiar properties of cold atomic systems will soon create new out-of-equilibrium paradigms,
where interactions and unitary dynamics are relevant; these studies will pave the way for their understanding, 
well beyond the commonly employed linear-response approach \cite{Thermal}.

\paragraph{Acknowledgements. ---}
We are indebted to R. Fazio for enlightening comments and support.
We also acknowledge fruitful discussions 
with D. Bernard, P. Calabrese, B. Doyon, F. Essler, J. Moore and in particular A. Kl\"umper. 
This work was supported by Italian MIUR
via FIRB Project RBFR12NLNA, and by Regione Toscana POR FSE 2007-2013. 
A.D.L. thanks Scuola Normale Superiore for hospitality.

\AtEndEnvironment{thebibliography}{
\bibitem{Thermal}  
  Ch. Greiner, C. Kollath, and A. Georges, arXiv:1209.3942,  arXiv:1406.4632;
  H. Kim and D. A. Huse,  Phys. Rev. A {\bf 86}, 053607 (2012);
  E. L. Hazlett, L.-C. Ha, and C. Chin,  arXiv:1306.4018;
  A. Rancon, C. Chin, and K. Levin,  arXiv:1311.0769;
  G. Benenti, G. Casati, T. Prosen, and K. Saito,  arXiv:1311.4430 (2013);
  C.-C. Chien, M. Di Ventra and M. Zwolak, arXiv:1403.0511 (2014).
  }

\bibliography{noneq}

\clearpage
\newpage

\clearpage 
\setcounter{equation}{0}%
\setcounter{figure}{0}%
\setcounter{table}{0}%
\renewcommand{\thetable}{S\arabic{table}}
\renewcommand{\theequation}{S\arabic{equation}}
\renewcommand{\thefigure}{S\arabic{figure}}

\onecolumngrid

\begin{center}
{\Large Supplementary Material for EPAPS \\ 
\titleinfo
}
\end{center}

\section{Quantum transfer matrix\label{sec_QTM} and Generalized Gibbs Ensemble} 
\noindent
Here, we discuss how our Ansatz for the stationary state out-of-equilibrium can be interpreted as GGE.
Let us consider the following non-normalized density matrix
\beq
\label{general_rho}
\rho=e^{-\sum_{\mm=1}^{\infty}\lambda_{\mm} \hat{\mathcal{Q}}_{\mm}},
\eeq
 where $\hat{\mathcal{Q}}_{\mm}$ are mutually commuting operators defined in the text by means of the boost operator and the parameters $\lambda_{\mm}\in\mathbb R$ are suitably chosen Lagrange multipliers.
 Given the quantum state \eqref{general_rho}, the QTM formalism introduced
 in \cite{klumper1992free, klumper1993thermodynamics} and recently considered in  \cite{pozsgay2013generalized, fagotti2013stationary},
 allows computing the generalized free-energy
 \beq
 \label{general_gf}
 \phi_{\rho}(\{\lambda\})\equiv\log\Tr[e^{-\sum_{\mm=1}^{\infty}\lambda_\mm \hat{\mathcal{Q}}_\mm}].
 \eeq
In the particle-hole setup \cite{klumper1993thermodynamics}, one has explicitly
 \beq
 \label{gf_ph}
 \phi_{\rho}(\{\lambda\})=-\sum_{\mm=1}^{\infty}\lambda_\mm f_\mm+\int_{-\infty}^{\infty} d\uu~\frac{\log\bigl[B(\uu)\bar{B}(\uu)\bigr]}{4\cosh (\pi \uu/2)},
 \eeq
 where  $B=1+b$ and $\bar{B}=1+\bar{b}$. The functions $b, \bar{b}$ are the solutions of the non-linear integral equation \eqref{NLIE} setting
 \beq
 \label{sourcegeneral}
  s(\uu) = - \sum_{\mm=1}^\infty \lambda_\mm \varepsilon_\mm(\uu)
 \eeq
and the constants $f_{\mm}$ are the expectation values of the charges $\mathcal{Q}_{\mm}$ in the ground-state, given by
\beq
\label{gscharges}
f_\mm = \pi \gamma \Lambda^\mm \int_{-\infty }^{\infty } d\omega  (i\omega)^{\mm-1} \left[\frac{\tanh \gamma  \omega}{\tanh \pi \omega}-1\right]\;.
\eeq
The QTM formalism can be used to obtain expectation values for all the
conserved charges inside
the  state \eqref{general_rho}. For instance for $\hat{\mathcal{Q}}_{\mm}$, we modify the free-energy as
\beq
\label{general_gf_mu}
\phi_{\rho}(\{\lambda\},\mu)\equiv\log\Tr[\rho e^{\mu\hat{\mathcal Q}_{\mm}}]\;
\eeq
and it can be obtained shifting the $k$-th Lagrange multiplier $\lambda_{\mm} \to \lambda_{\mm} - \mu$.
Finally the expectation value of the charge density $q_{\mm}$ is given by the derivative
\beq
\label{qder}
q_{\mm} = \left.\partial_{\mu}\phi_{\rho}(\{\lambda\},\mu)\right|_{\mu=0}.
\eeq
In principle, this value can be obtained computing \eqref{general_gf_mu} numerically solving \eqref{NLIE} for different $\mu$ and then using \eqref{qder}. However,
a numerically more stable procedure is to explicitly differentiate $\phi_{\rho}(\{\lambda\},\mu)$ with respect to $\mu$ directly from its expression \eqref{gf_ph}.
One obtains \eqref{charges} where 
\beq
\label{adef}
a_{\mm} = \left.\partial_\mu \log b(\uu)\right|_{\mu=0}\;.
\eeq
This last quantity solves \eqref{linearak}, which is derived differentiating 
\eqref{NLIE} with respect to $\mu$ at $\mu = 0$.

\vspace{0.5cm}
\noindent In our approach the Lagrange multipliers in \eqref{general_rho} are implicitly fixed by the source term, that we choose to be
\beq
s(\uu)=-[\beta_l\theta(\uu)+\beta_r\theta(-\uu)\bigr] \varepsilon(\uu) \;.
\eeq
To show that this can be written in the form \eqref{general_rho}, we look for a set of $\lambda$'s solving the equation
\beq
\label{lagrange_mult}
\sum_{\mm=1}^\infty \lambda_{\mm}\varepsilon_{\mm} (\uu) = -s(\uu)\;.
\eeq
Denoting $\tilde f(q)$ the Fourier transform of $f(\uu)$
\beq
\label{Fourier}
f(\uu)=\int_{-\infty}^{\infty} \frac{dq}{2\pi}~ e^{iqx} \tilde{f}(q),
\eeq
and using $\varepsilon_{\mm+1}(\uu)= \Lambda^{\mm}
\varepsilon^{(\mm)}(\uu)$, we have
\beq
\label{epshat}
\tilde \varepsilon_{\mm}(q) = \left(i q \Lambda\right)^{\mm-1} \tilde\varepsilon(q)
\eeq
and \eqref{lagrange_mult} is then solved by
\beq
\label{lagrange_sol}
\lambda_{\mm+1}=-\left(\frac{1}{i\Lambda}\right)^{\mm}\frac{1}{\mm!}\left.\frac{d^\mm}{dq^{\mm}}\frac{\tilde{s}(q)}{\tilde{\varepsilon}(q)}\right|_{q=0}\;.
\eeq
Notice that all the derivatives are well defined since $\int_{-\infty}^\infty d\uu \uu^k \varepsilon(\uu) <\infty$ and 
the parameters $\lambda_{\mm}$ are fixed uniquely. Convergence of the LHS of
\eqref{lagrange_mult} is pointwise with the  exception of the point $\uu=0$. 

\section{Conformal limit \label{app_conflim}}
\noindent In this Appendix we derive the results (\ref{factorization},\ref{charge_CFT}).
Let us consider the functions $b_{+}(\xi)=b\bigl(\uu_++\frac{2}{\pi}\xi\bigr)$ and $b_{-}(\xi)=b\bigl(-\frac{2}{\pi}\xi\bigr)$ for positive $\xi$ and zero otherwise. The same definitions are used for the bar-quantities.
For large $\beta_{l}$, $s\bigl(\uu_+ + \frac{2}{\pi}\xi\bigr) \simeq -e^{-\xi}$ and  \eqref{NLIE} can be reduced to  
\begin{align}
\label{NLIECFTplus}
\left[\begin{array}{c}
\log b_{+}(\xi)\\
\log\bar{b}_{+}(\xi)
\end{array} \right]&=-\left[\begin{array}{c}
e^{-\xi}\\
e^{-\xi}
\end{array}\right]+
\left[\begin{array}{cc}
\tilde{\kk}& -\tilde{\kk}_{\mp}\\
-\tilde{\kk}_{\pm}& \tilde{\kk}
\end{array}\right]\star\left[\begin{array}{c}
\log B_{+}(\xi)\\
\log\bar{B}_{+}(\xi)
\end{array}\right],
\\
\label{NLIECFTminus}
  \left[\begin{array}{c}
      \log b_-(\xi)\\
      \log\bar{b}_-(\xi)
    \end{array}\right]&=\beta_r \left[\begin{array}{c}
      \varepsilon(\frac{2\xi}{\pi})\\
      \varepsilon(\frac{2\xi}{\pi})
    \end{array}\right]+\left[\begin{array}{cc}
      \tilde\kk& -\tilde\kk_+ \\
      -\tilde\kk_-& \tilde\kk
    \end{array}\right]\star\left[\begin{array}{c}
      \log B_{-}\\
      \log \bar{B}_{-}
    \end{array}\right],
\end{align}
where $B_{\pm}(\xi) = 1+b_{\pm}(\xi)$ and the kernels
$\tilde{\kk}(\xi)=\frac{2}{\pi}\kk\bigl(\frac{2\xi}{\pi}\bigr)$
and $\tilde{\kk}_{\pm}(\xi)=\frac{2}{\pi}\kk\bigl(\frac{2\xi}{\pi}\pm 2i\bigr)$. Notice that $\tilde{\kk}(\xi)=\tilde{\kk}(-\xi)$. 
Moreover we set $a_{\mm}^+(\xi)=a_{\mm}\bigl(\uu_++\frac{2}{\pi}\xi\bigr)$ and $a_{\mm}^-(\xi)=a_{\mm}\bigl(-\frac{2}{\pi}\xi\bigr)$. Eq.~\eqref{linearak} simplifies for large $\beta_{l}$ to 
\begin{align}
\label{linearCFTplus}
  \left[\begin{array}{c}
a_{\mm}^{+}(\xi)\\
\bar{a}_{\mm}^{+}(\xi)
\end{array}\right]&=\frac{1}{\beta_l}\left(-\frac{\pi\Lambda}{2}\right)^{\mm-1}\left[\begin{array}{c}
e^{-\xi}\\
e^{-\xi}
\end{array}\right] +\left[\begin{array}{cc}
\tilde{\kk}& -\tilde{\kk}_{\mp} \\
-\tilde{\kk}_{\pm}& \tilde{\kk}
\end{array}\right]\star\left[\begin{array}{c}
n_{+} a_{\mm}^{+}(\xi)\\
\bar{n}_{+}\bar{a}_{\mm}^{+}(\xi)
\end{array}\right],
\\
\label{linearCFTminus}
  \left[\begin{array}{c}
a_{\mm}^{-}(\xi)\\
\bar{a}_{\mm}^{-}(\xi)
\end{array}\right]&=\left[\begin{array}{c}
      \varepsilon_m(-\frac{2\xi}{\pi})\\
      \varepsilon_m(\frac{2\xi}{\pi})
    \end{array}\right]
    +\left[\begin{array}{cc}
\tilde{\kk}& -\tilde{\kk}_{\mp} \\
-\tilde{\kk}_{\pm}& \tilde{\kk}
\end{array}\right]\star\left[\begin{array}{c}
n_{-} a_{\mm}^{-}(\xi)\\
\bar{n}_{-}\bar{a}_{\mm}^{-}(\xi)
\end{array}\right].
\end{align}
where $n_{\pm} = b_{\pm}/(1+b_{\pm})$. Combining (\ref{NLIECFTplus}, \ref{linearCFTplus}), one  realizes that
\beq
\label{derivativesa}
a_{\mm}^{+}(\xi)=\frac{1}{\beta_l}\left(-\frac{\pi\Lambda}{2}\right)^{\mm-1}\frac{d}{d\xi}\log b_+(\xi).
\eeq
Rewriting \eqref{charges} with the functions $n_{\pm}$ and $a_{\pm}$, we deduce the decomposition
\beq
\label{decomposition}
q_{\mm} = f_{\mm} + \underbrace{\frac{2}{\pi}\int_{0}^\infty \frac{n_-(\xi) a_m^-(\xi)}{4 \cosh \xi }d\xi}_{q^-_{\mm}(\beta_r)}  + \underbrace{\frac{1}{\pi^2 \Lambda \beta_l}\int_{-\infty}^\infty e^{-\xi}\;	 n_+(\xi) a_m^+(\xi) d\xi}_{q^+_{\mm}(\beta_l)	} + \ldots
\eeq
where for simplicity we omit the bar-terms that have a similar decomposition, thus recovering the factorized form \eqref{factorization}. 
The expression for $q_{\mm}^{+}$ in \eqref{factorization} in the large-$\beta_{l}$ limit can be integrated by parts with the aid of \eqref{derivativesa} to get 
\beq
\label{qCCFT1}
q_{\mm}^{+}(\beta_l)\sim-\frac{c_{\mm}}{2\pi\beta_l^2}\int_{\mathbb R}d\xi e^{-\xi}\log[B_+(\xi)\bar{B}_+(\xi)],
\eeq
where $c_{\mm+2}=\bigl(-\pi\Lambda/2\bigr)^{\mm}$. The integral in \eqref{qCCFT1} can be computed exactly \cite{klumper2002thermal}. Indeed, taking the
$\xi$-derivative in \eqref{NLIECFTplus},  multiplying the result on the left by $[\log B_{+}, \log\bar{B}_{+}]$ and finally
integrating one obtains
\beq
\label{dilog}
\int_{\mathbb R}d\xi~e^{-\xi}
\log\bigl[B_{+}(\xi)\bar{B}_{+}(\xi)\bigr]=2\int_{b_+(-\infty)}^{b_{+}(\infty)}\frac{db}{b}\log(1+b).
\eeq
Observing that $b_+(-\infty)=0$ and $b_{+}(\infty)=1$ and recalling
$\int_{0}^{1}d\uu\log(1+\uu)\uu^{-1}=\pi^2/12$,  the result \eqref{charge_CFT} now follows. 
\section{High Temperature limit \label{app_hight}}
We parameterize $\beta_r=\eta\beta_l=\beta$ and consider the limit $\beta\rightarrow 0$ where 
$b=1+\beta\partial_{\beta} b|_{\beta=0}+o(\beta)$ and similarly for $\bar{b}$; then linearizing \eqref{NLIE} we obtain the following integral
equation for the auxiliary function derivatives at $\beta=0$
\begin{align}
  \label{L_b}
  \left[\begin{array}{c}
      \partial_{\beta} b|_{\beta=0}\\
      \partial_{\beta}\bar{b}|_{\beta=0}
    \end{array}\right]=-\frac{1}{2}\left[\begin{array}{c}
      \varepsilon(u)[\eta+(1-\eta)\theta(-u)]\\
      \varepsilon(u)[\eta+(1-\eta)\theta(-u)]
    \end{array}\right]+\frac{1}{2}\left[\begin{array}{cc}
      \kk& -\kk_+ \\
      -\kk_-& \kk
    \end{array}\right]\star\left[\begin{array}{c}
      \partial_{\beta} b|_{\beta=0}\\
      \partial_{\beta}\bar{b}|_{\beta=0}
    \end{array}\right].
\end{align}
The current slope  at high temperatures can be determined from \eqref{charges} expanding the integrand at order $\beta$; one has
$\mathcal{J}(\beta)=\beta \kappa+o(\beta)$ with
\beq
\label{slope}
\kappa=\int_{-\infty}^{\infty}\frac{d\uu}{16\cosh (\pi \uu/2)}~\left[\partial_\beta b|_{\beta=0}a_{0}+2\partial_{\beta}a|_{\beta=0}+\partial_\beta \bar{b}|_{\beta=0}\bar{a}_{0}+2\partial_{\beta}\bar{a}|_{\beta=0}\right],
\eeq
where $a(\beta)=a_{0}+\beta\partial_{\beta}a|_{\beta=0}+o(\beta)$ and analogously for $\bar{a}$. Linear integral equations for $a_0, \bar{a}_0$ and
$\partial_{\beta}a|_{\beta=0},\partial_{\beta}\bar{a}|_{\beta=0}$ are derived from \eqref{linearak}, they read
\begin{align}
  \label{L_a}
  &\left[\begin{array}{c}
      a_0\\
      \bar{a}_0
    \end{array}\right]=\left[\begin{array}{c}
     \varepsilon_2\\
     \varepsilon_2
    \end{array}\right]+\frac{1}{2}\left[\begin{array}{cc}
      \kk& -\kk_+ \\
      -\kk_-& \kk
    \end{array}\right]\star\left[\begin{array}{c}
      a_0\\
      \bar{a}_0
    \end{array}\right],\\
  \label{L_da}
  &\left[\begin{array}{c}
      \partial_{\beta} a|_{\beta=0}\\
      \partial_{\beta} \bar{a}|_{\beta=0}
    \end{array}\right]=\frac{1}{4}\left[\begin{array}{cc}
      \kk& -\kk_+ \\
      -\kk_-& \kk
    \end{array}\right]\star\left[\begin{array}{c}
     \partial_\beta b|_{\beta=0}a_{0}+2\partial_{\beta}a|_{\beta=0} \\
     \partial_\beta \bar{b}|_{\beta=0}\bar{a}_{0}+2\partial_{\beta}\bar{a}|_{\beta=0} 
    \end{array}\right].  
\end{align}
Eq. \eqref{L_a} can be solved analytically \cite{klumper2002thermal}, whereas \eqref{L_b} and \eqref{L_da} only numerically. We finally set $\eta=1/2$ and employed
\eqref{slope} to reproduce the linear behavior of the current plotted  in the inset of Fig. \ref{fig:Figure4} .

\end{document}